\documentclass[aps,twocolumn,showpacs,float]{revtex4}

\usepackage{amsmath}
\usepackage{graphicx}
\usepackage{float}
\floatstyle{boxed}
\usepackage{bm}

\begin{document}

\bibliographystyle{prsty}
\author{Liufei Cai$^{1}$, Reem Jaafar$^2$, and Eugene M. Chudnovsky$^1$}
\affiliation{$^{1}$Physics Department, Lehman College, The City University of New
York, 250 Bedford Park Boulevard West, Bronx, NY 10468-1589\\
$^{2}$Department of Mathematics, Engineering and Computer Science, LaGuardia Community College, The City University of New York, 31-10 Thomson Avenue, Long Island City, NY 11101}
\date{\today}

\begin{abstract}
Switching of the direction of the magnetic moment in a nanomagnet is studied within a modified Slonczewski's model that permits torsional oscillations of the magnet. We show that the latter may inhibit or assist the magnetization switching, depending on parameters. Three regimes have been studied: the switching by torsional oscillations alone, the switching by the spin-polarized current with torsional oscillations permitted, and the magnetization switching by the current combined with the mechanical twist. We show that switching of the magnetic moment is possible in all three cases and that allowing torsional oscillations of the magnet may have certain advantages for applications. Phase diagrams are computed that show the range of parameters required for the switching.  
\end{abstract}

\title{Mechanically-Assisted Current-Induced Switching of the Magnetic Moment in a Torsional Oscillator}

\maketitle

\section{Introduction}

The interest in spin dynamics operated by the electric current started with the suggestion \cite{Bergeretal} to move domain walls in metals by the flow of spin-polarized electrons \cite{Parkin}. The discovery of the interlayer exchange coupling and giant magnetoresistance in magnetic multilayers \cite{Fert,Grunberg} gave further boost to research on current-induced switching of magnetization. Slonczewski \cite{Slonczewski96} and Berger \cite{Berger96} demonstrated that a spin-polarized current could deliver a spin transfer torque that is sufficient to reorient the magnetic moment in a nanostructure. This suggestion triggered a wide-spread research on magnetic devices operated by spin-polarized currents \cite{Ralph}. The field has progressed towards current-induced magnetization switching in the smallest nanostructures, including nanowires. 

The question we are addressing in this paper is whether some degree of the mechanical freedom of the nanostructure would assist or inhibit the process of the magnetization switching. This question may be of particular importance because electric currents needed to switch the magnetization are rather high. Meanwhile, as is well known, the magnetization can be switched by the low amplitude high-frequency ac magnetic field. When absorbed in a resonant manner, it drives the magnetic moment up the energy barrier towards the reversal \cite{CGC-PRB2013}. The corresponding wavelengths of the electromagnetic field are typically in the centimeter range, making it difficult to selectively apply this method to small densely packed memory units. This prompted some researchers to use Josephson junctions for generation of the ac field in a small magnet \cite{Wern-SST2009}. The latter method, however, requires low temperatures. 

Our work is motivated by the observation that, in the coordinate frame coupled to the magnetic anisotropy axes, the rotational vibrations of a magnet are equivalent to the ac magnetic field \cite{CGS-PRB2005}. Consequently, for nanomechanical resonators with vibrational frequencies comparable to the frequency of the ferromagnetic resonance, the effect of rotational vibrations can be similar to the effect of the ac magnetic field. Unlike the ac field, however, the mechanical vibrations can be localized at the nanoscale. 

The connection between the magnetization reversal and mechanical rotation has been known for almost a century. It is manifested in the Barnett and Einstein - de Haas effects \cite{Barnett,EdH}. Coupling of mechanical resonators to classical magnetic moments has been studied in the past in the context of magnetization reversal in a thin magnetic film deposited on a microcantilever \cite{Kovalev-PRL2005} and nanomachines operated by the ac currents \cite{Stiles-JAP2002,Mohanty-PRB2004,Kovalev-PRB2007}. Einstein - de Haas effect in a magnetic microcantilever has been measured \cite{Wallis-APL2006} and explained \cite{JCG-PRB2009} by the motion of a domain wall. Various aspects of the electronic transport through magnetic molecules have been investigated together with the possibility of writing, storing, and reading spin information in memory devices based upon single-molecule magnets \cite{Timm-PRB2006}. Switching of the molecular spin by a spin-polarized current through a molecule bridged between conducting leads has been proposed \cite{Misiorny-PRB2007} and the effect of a soft vibrating mode of the molecule on the electronic transport has been studied \cite{Cornaglia-PRB2007}. Experiment has progressed to the measurement of the spin reversal in a single-molecule magnet drafted on a carbon nanotube \cite{Wern-NataureNano2013,Wern-ASC-Nano2013}. 

In this paper, we extend the model of Slonczewski \cite{Slonczewski96} for a situation when the magnet subjected to a spin-polarized current is free to develop torsional oscillations. Our focus is on the magnetization reversal by the dc current, assisted by mechanical vibrations. Dynamics of the total angular momentum, spin plus mechanical angular momentum, makes this problem uniquely defined and free of any unknown coupling constants \cite{EC-Springer}. The paper is structured as follows. The model is formulated in Section \ref{formulation}. Mechanically assisted magnetization switching (MAMS) in the absence of the current is studied in Section \ref{mechanical}. Switching by the current in a system that permits torsional oscillations is investigated in Section \ref{current}. Section \ref{combined} deals with the magnetization switching assisted by both, the spin-polarized current and independently generated torsional oscillations. Results and possible implementations of the proposed mechanisms of magnetization switching are discussed in Section \ref{discussion}. 
 
\section{Formulation of the Problem} \label{formulation}

In the original formulation of the problem by Slonczewski the magnetic moment ${\bf m}$ of an uniaxial nanomagnet is switched by a spin-polarized current $I$ originating from a ferromagnet that has a fixed direction of the magnetization ${\bf M}$. The dynamics of ${\bf m}$ is governed by the modified Landau-Lifshitz equation \cite{Lectures} called the Slonczewski equation\cite{Slonczewski96,Ralph},
\begin{align}
\dot{\bf m}=\hat{\bf m}\times\left[-\gamma_gH_u({\bf c}\cdot{\bf m}){\bf c}+\alpha\dot{\bf m}+g(\theta)\frac{\mu_BI}{e}\hat{\bf m}\times\hat{\bf M}\right]
\label{slonczewski-eq}
\end{align}
Here $H_u$ is the magnitude of the uniaxial magnetic anisotropy field, $\hat{\bf m}$ and $\hat{\bf M}$ are unit vectors along ${\bf m}$ and ${\bf M}$, and ${\bf c}$ is the unit vector along the easy magnetization axis. Parameters $\gamma_g$, $\alpha$, $\mu_B$ , and $e$ are respectively the gyromagnetic ratio, the Gilbert damping, the Bohr magneton, and the electron charge. Function $g(\theta)$ is given by \cite{Slonczewski96} $g(\theta)=\left[-4+{(1+P)^3(3+cos\theta)}/({4P^{3/2}})\right]^{-1}$ where $\theta$ is the angle between ${\bf m}$ and ${\bf M}$ and $P$ is the spin polarization of the ferromagnetic material, $0 < P < 1$. 

Using spherical angles for the magnetic moment ${\bf m}$ that undergoes the reversal, one can obtain from Eq.\ (\ref{slonczewski-eq})
\begin{align}
\dot{\theta}=\left[g(\theta)\frac{\mu_BI}{em}-\alpha \gamma_g H_ucos\theta\right]sin\theta
\label{slonczewski-theta}
\end{align}
where $m$ is the absolute value of ${\bf m}$.  

For complete switching one should require $g(\theta){\mu_BI}/({em})-\alpha \gamma_g H_ucos\theta>0$,
which yields the Slonczewski condition, $I>I_{sc}$, for the magnetization reversal, with the critical current $I_{sc}$ given by $I_{sc}=k(P) \alpha (m/\mu_B) e\omega_{FMR}$. Here $\omega_{FMR}=\gamma_gH_u$ is the frequency of the ferromagnetic resonance and $k(P)$ is the numerical factor of order unity that depends on the spin polarization of the current. For iron $P=0.4$ while some materials may have $P$ close to one. If one writes $g(\theta)=(A-B cos\theta)^{-1}$ then $k(P) = A^2/(4B)$. For the numerical work we are choosing $P=0.5$, which according to Eq.\ (\ref{slonczewski-theta}) corresponds to $A = 3.16$, $B = 2.39$, $k = 1.04$.

\begin{figure}
\includegraphics[width=80mm]{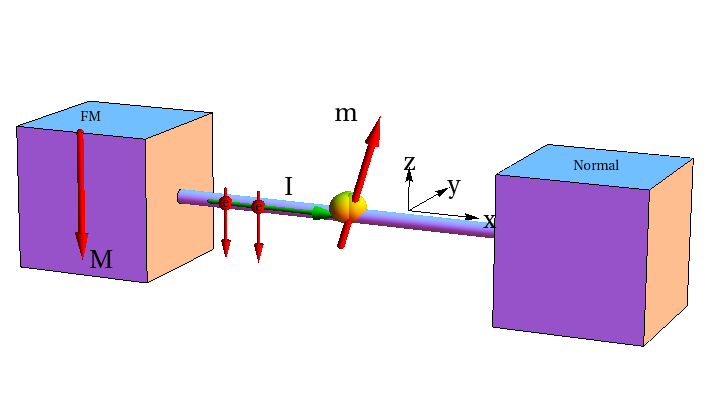}
\caption{Schematic presentation of the geometry of the model studied in the paper (actual dimensions may vary). Nanomagnet is a part of a torsional resonator. Spin tranfer torque delivered by the spin-polarized electric current causes rotation of the magnetic moment that is coupled to the mechanical rotation of the resonator through Einstein - de Haass effect.}
\label{magnetic-molecule-model}
\end{figure}
The geometry of our model (not the actual proportions) is depicted in Figure.\ref{magnetic-molecule-model}. It is conceptually similar to the Slonczewski's model with the only difference that the nanomagnet is now a part of a torsional oscillator. Consequently, the direction of its easy magnetization axis ${\bf c}$ is allowed oscillate in space in accordance with the coupled dynamics of the magnetic moment and the oscillator. The axis of the mechanical rotation is along the $\hat{\bf x}$ direction. When the torsional oscillator rotates by the angle $\rho$ about the $\hat{\bf x}$ axis, the direction of the easy axis ${\bf c}$ in Eq.\ (\ref{slonczewski-eq})  transforms as ${\bf c}(\rho)=R(\rho){\bf c}$ were $R(\rho)$ is the rotation matrix 
\begin{equation}
R(\rho)= \left(\begin{array}{ccc}
          1 & 0 & 0\\
          0 & cos\ \rho & -sin\ \rho\\
          0 & sin\ \rho & cos\ \rho
\end{array} \right) 
\end{equation}
 We choose the direction of the magnetization of the ferromagnetic source of the spin polarized current, ${\bf M}$, to be along the $-\hat{\bf z}$ direction and the equilibrium orientation of the easy axis ${\bf c}$ of the nanomagnet to be along the $\hat{\bf z}$ direction. This gives 
\begin{eqnarray}\label{c}
&& {\bf c}(\rho)=(0,-sin\rho,cos\rho)\\
&&{\bf c}(\rho)\cdot{\bf m}=m_z cos\rho-m_y sin\rho
\end{eqnarray}
Substituting this into Eq.\ (\ref{slonczewski-eq}) and writing ${\bf m}=m(sin\theta cos\phi,sin\theta sin\phi,cos\theta)$ we obtain
\begin{eqnarray}
&&\frac{d\theta}{dt'}=\left[g(\theta)I'-\alpha\dot{\phi}\right]sin\theta  \label{dottheta} \\
&&-sin\rho \,cos\phi(cos\theta \,cos\rho-sin\rho \,sin\theta \,sin\phi)  \nonumber \\
&&\frac{d\phi}{dt'}=\frac{\alpha g(\theta)I'}{1+\alpha^2}+\frac{1}{1+\alpha^2}\left(cos\rho+sin\rho \,cot\theta \,sin\phi \right.  \label{dotphi} \\
&&-\left.\alpha {sin\rho}\,cos\phi/{sin\theta}\right)(cos\theta \,cos\rho-sin\rho \,sin\theta\, sin\phi) \nonumber
\end{eqnarray}
where we switched to dimensionless 
\begin{equation}
t'=\omega_{FMR}t, \quad I'=\frac{\mu_B}{m}\frac{I}{e\omega_{FMR}}, \quad {\bf m}'=\frac{\bf m}{m}
\end{equation} 

The above equations for ${\bf m}$ must be accompanied by the equation of motion for the torsional oscillator. The latter follows from the equation $\dot{J}_{x}={\tau}_{x}$ where ${\bf J}$ is the total angular momentum and ${\bm \tau}$ is the torque. Their $x$-components are given by 
\begin{eqnarray}
&&{J}_{x}=-\frac{m_x}{\gamma_g}+I_r\dot\rho \label{momentum}\\
&& {\tau}_{x}=-g(\theta)\frac{\mu_B I}{e\gamma_g}\hat{\bf m}\times(\hat{\bf M}\times\hat{\bf m})\cdot\hat{\bf x}-I_r\omega_0^2\rho \label{torque}
\end{eqnarray}
respectively, where $I_r$ is the moment of inertia of the rotator and $\omega_0$ is the frequency of its torsional vibrations. The first term in Eq.\ (\ref{momentum}) is the spin angular momentum and the second term is the mechanical angular momentum. Similarly, the first term in Eq.\ (\ref{torque}) is the spin transfer torque from polarized electrons and the second term is the returning mechanical torque due to the elastic twist. Here we are assuming that the latter is proportional to the angle of twist $\rho$. Note that, in, e.g., nanowires this assumption is justified even for large rotation angles \cite{Weinberger-Nano2009}. The negative sign of ${\bf m}$-terms is due to the fact that electron spin is opposite to the direction of the magnetization ($e$ and $\gamma_g$ are considered to be positive constants.). Introducing the dimensionless damping parameter of the torsional oscillator $\eta$ we obtain from $\dot{J}_{x}={\tau}_{x}$
\begin{align}
\frac{d^2\rho}{dt'^2} +\omega_0'\eta \frac{d\rho}{dt'}+\omega_0'^2\rho=\frac{1}{I_r'}\frac{d{m}_x'}{dt'}-\frac{g(\theta)I'}{I_r'}m_z'm_x'
\label{rho-eq}
\end{align}
Here we use dimensionless parameters 
\begin{equation}
\omega_0'=\frac{\omega_0}{\omega_{FMR}}, \quad I_r'=\frac{\gamma_g\omega_{FMR}}{m}I_r
\end{equation}
Full dynamics of the system is described by equations (\ref{dottheta}), (\ref{dotphi}), and (\ref{rho-eq}). 
Note that when $\rho=0$, equations (\ref{dottheta}) and (\ref{dotphi}) reduce to Eq.\ (\ref{slonczewski-theta}). 

\section{Magnetization Switching by Torsional Oscillations in the Absence of the Current} \label{mechanical}

\begin{figure}
\includegraphics[width=70mm]{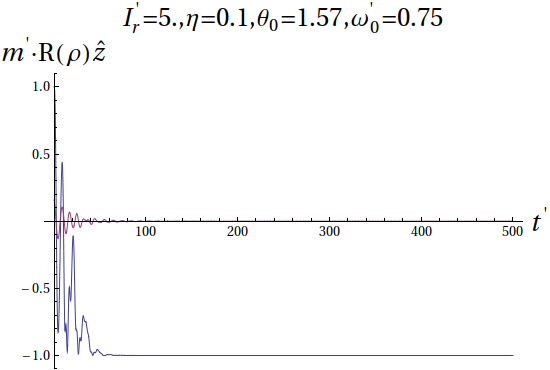}{a}
\vspace{0.1in}
\includegraphics[width=70mm]{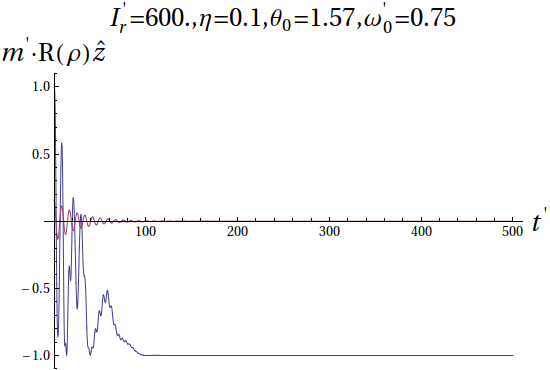}{b}
\vspace{0.1in}
\includegraphics[width=70mm]{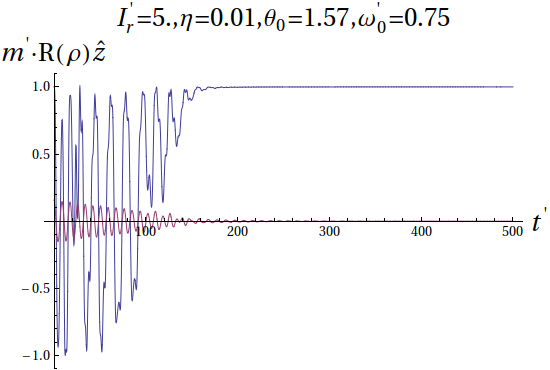}{c}
\vspace{0.1in}
\includegraphics[width=70mm]{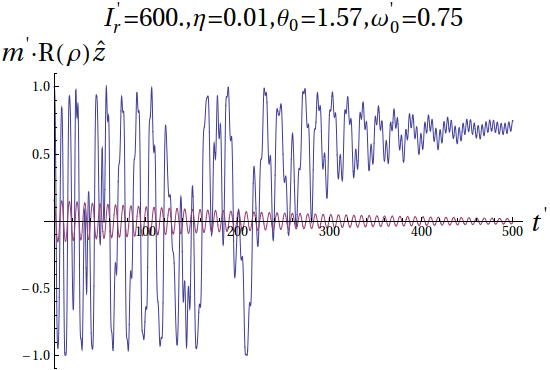}{d}
\caption{Dynamics of the magnetic moment induced by the $\pi/2$ mechanical twist of the torsional oscillator in the absence of spin polarized current for two moments of inertia, $I' = 5$ and $I' = 600$. Blue line shows oscillations of the magnetic moment projected onto the magnetic anisotropy axis, ${\bf m}' \cdot R(\rho)\hat{\bf z}$. Red line shows mechanical oscillations $\rho(t')$. (a,b) Magnetization witching with damping $\eta=0.1$; (c,d) No switching with damping $\eta=0.01$.}
\label{switching-no-current}
\end{figure}
Numerical solution of equations (\ref{dottheta}),  (\ref{dotphi}), and (\ref{rho-eq}) at $I=0$ reveals that the magnetic moment can be switched by the vibrational motion of the torsional oscillator alone if the amplitude of the oscillations is sufficiently large.  We call such a process the Mechanically-Assisted Magnetization Switching (MAMS). To initiate MAMS one twists the oscillator by a large angle, e.g., $\rho_0= \pi/2$, waits until ${\bf m}$ comes to thermal equilibrium by aligning with the new direction of the anisotropy axis ${\bf c}$, and then releases the oscillator. The initial condition is, therefore, $\theta_0 = \rho_0$.  We find that MAMS is not very sensitive to the moment of inertia of the torsional oscillator but is sensitive to its resonance frequency, its damping constant, and the angle of twist, see Fig.\ \ref{switching-no-current}. For every set of parameters the switching occurs in a certain range of the mechanical damping $\eta$. It does not occur for very small $\eta$ or very large $\eta$.

\begin{figure}
\includegraphics[width=60mm]{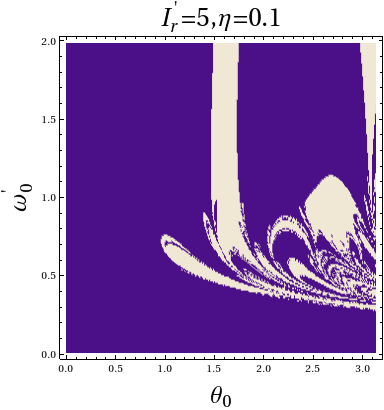}{a}
\includegraphics[width=60mm]{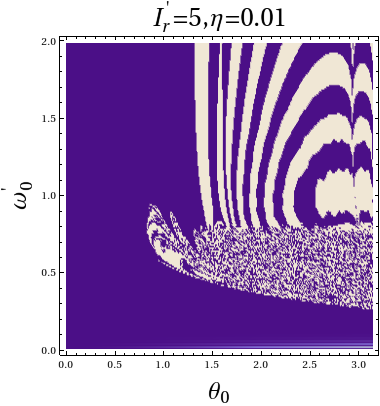}{b}
\caption{ MAMS phase diagram, $\omega_0'$ vs $\theta_0$, for $I_r' = 5$ and two different values of $\eta$. The area of switching is shown by light color.}
\label{phasediagram-mams-I5}
\end{figure}
\begin{figure}
\includegraphics[width=60mm]{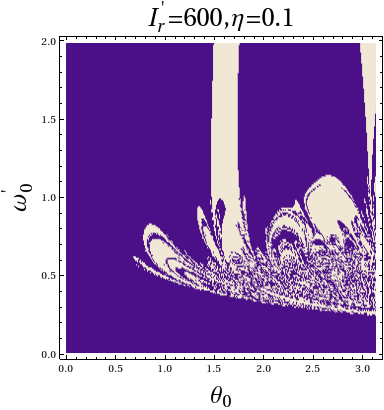}{a}
\includegraphics[width=60mm]{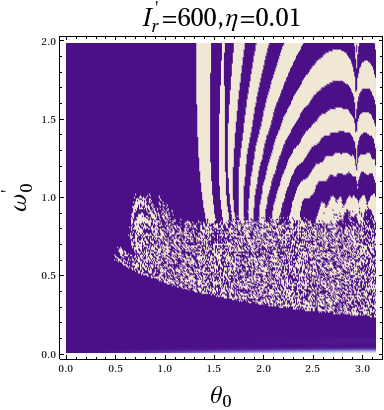}{b}
\caption{ MAMS phase diagram, $\omega_0'$ vs $\theta_0$, for $I_r' = 600$ and two different values of $\eta$. The area of switching is shown by light color.}
\label{phasediagram-mams-I600}
\end{figure}
\begin{figure}[ht]
\includegraphics[width=70mm]{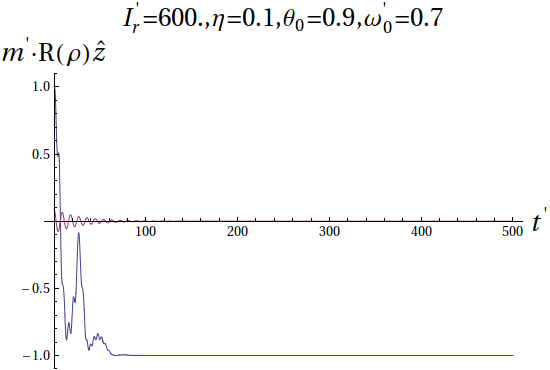}
\caption{ Dynamics of switching at a point $\omega_0'=0.7$ and $\theta_0=0.9$ in the phase diagram shown in Fig.\ref{phasediagram-mams-I600}(a).}
\label{mams}
\end{figure}

Switching phase diagrams of $\omega_0'$ vs $\theta_0$, for $I_r'=5$ and $I_r'=600$, $\eta=0.1$ and $\eta=0.01$, are shown in Figs. \ref{phasediagram-mams-I5} and \ref{phasediagram-mams-I600}. Switching areas are depicted in light color. There are continuous areas of switching as well as areas of sporadic switching where the dynamics of the system is extremely sensitive to the parameters. It is interesting to notice that at $\omega_0'<1$ a higher mechanical damping, $\eta=0.1$, gives cleaner and faster switching than lower damping, $\eta=0.01$. This is also clear from Fig. \ref{switching-no-current}. To confirm and illustrate this observation we plot in Fig.\ \ref{mams} the dynamics corresponding to the point of $(0.9,0.7)$ in Fig.\ \ref{phasediagram-mams-I600}(a).

The initial angle of twist $\rho_0 = \theta_0 = \pi/2$ is rather large and one might look for alternative ways to induce MAMS. One such way is to let the angle of twist oscillate for a certain period of time (say, from $t'=0$ to $t'=T_i$). This corresponds to replacing the dynamics described by Eq.\  (\ref{rho-eq}) with  $\rho(t)= \rho_0 \cos(t)$ and then, at  $t'=T_i$, letting Eq. (\ref{rho-eq}) govern the rest of the damped motion. In this case, for each set of parameters, there is a window of $\omega_0'$ for which the switching occurs and it is possible to induce the switching by smaller angles of twist, such as, e.g.,  $\pi/4$  and $\pi/3$. 

The above dynamics that includes driven oscillation followed by free oscillations produces well-defined areas as well as scattered areas of the switching in the phase diagram. For instance at $I_r'=5$, $\eta=0.1$, the switching is possible at $\rho_0=\theta_0=\pi/4$ for  $0.06 \leq  \omega_0' \leq 0.3  $ and also near $\omega_0'= 0.32$ and $\omega_0'=0.33$. If the damping $\eta$ is reduced to 0.01, the windows of switching are $0.07 \leq  \omega_0' \leq 0.35  $ and $0.38 \leq  \omega_0' \leq 0.48  $. At $\rho_0=\theta_0=\pi/3$, the are two switching windows for $0.01 \leq  \omega_0' \leq 0.12  $ and $0.21 \leq  \omega_0' \leq 0.60  $. One also finds narrow areas of switching near $\omega_0'= 0.14$ and $\omega_0'=0.62$. When the twist angle becomes very large (e.g. $\rho_0 = \pi/2$) the values of  $\omega_0'$ for which switching occurs become scattered. 

\begin{figure}
\includegraphics[width=70mm]{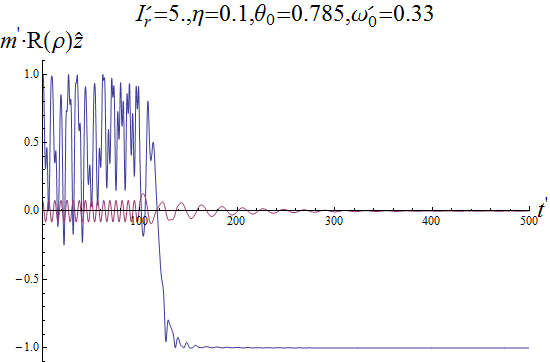}{a}
\includegraphics[width=70mm]{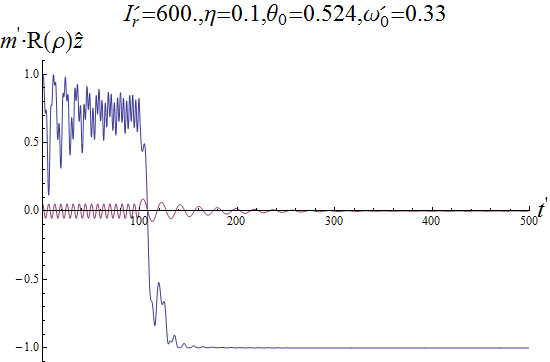}{b}
\caption{ Plots of $\rho(t)$ (red line) and magnetization in the coordinate frame of the mechanical oscillator, ${\bf m}'\cdot R(\rho)\hat{z}$ (blue line), for different values of parameters. Driven oscillations at $t' < 100$ are followed by free damped oscillations at $t' > 100$.}
\label{switchoscillatenoI}
\end{figure}

For a larger moment of inertia $I_r'=600$ it is possible to induce switching by the driven oscillations followed by free oscillations for angles smaller than $\pi/4$. For instance, if $\eta=0.1$ one can induce switching with $\rho_0=\theta_0=\pi/7$ in a narrow interval of resonance frequencies $0.29 \leq  \omega_0' \leq 0.36  $. This interval increases as $\rho_0$ increases but remains below $\omega_0'=0.5$ for angles of twist less than $\pi/3$. At $\rho_0 = \theta_0 =\pi/3$ the values of $\omega_0'$ for which switching occurs are scattered between  $0.01$ and $0.72$. Time dependence of the mechanical twist and the magnetic moment for two different sets of parameters is shown in Fig.\ \ref{switchoscillatenoI}. 

\section{Current-Induced Magnetization Switching in a Torsional Oscilator} \label{current}

\begin{figure}
\includegraphics[width=70mm]{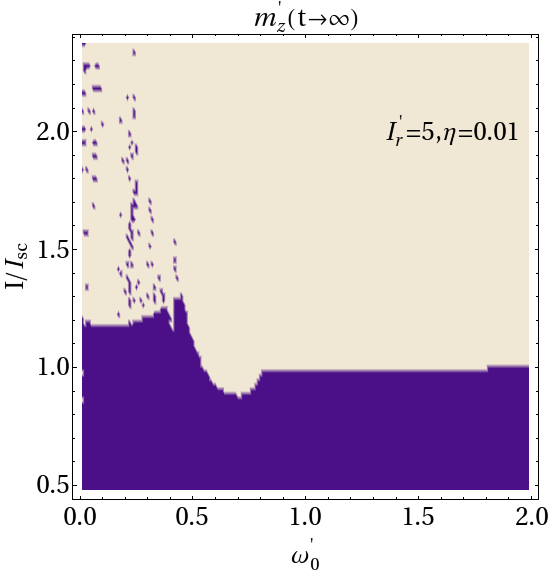}{a}
\includegraphics[width=70mm]{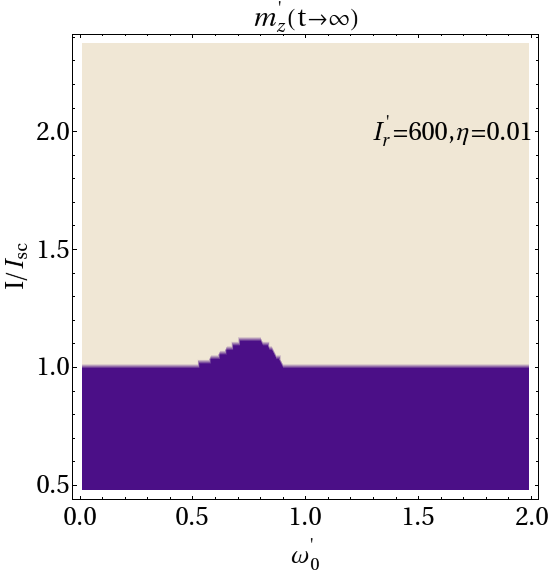}{b}
\caption{Switching phase diagrams for a current-driven magnetization reversal in a torsional oscillator. Switching area is shown in light color.}
\label{phasediagram-cisoc}
\end{figure}

In this Section we solve numerically equations (\ref{dottheta}),  (\ref{dotphi}), and (\ref{rho-eq}) at a non-zero current assuming that mechanical oscillation are generated by the current itself and not by any external force as in the previous Section. As is known \cite{Ralph}, to obtain a non-trivial dynajmics that leads to the magnetization reversal, one has to introduce small misalighnment of the equilibrium orientations of ${\bf M}$ and ${\bf m}$, that we set at $\theta_0 = 0.01$. Switching phase diagrams of $I/I_{sc}$ vs $\omega_0'$, are shown in Fig.\  \ref{phasediagram-cisoc} for $I_r'=5$ and $I_r'=600$ at $\eta = 0.01$. Switching areas are depicted in light color. The areas of switching are continuous but with many small islands embedded, where the switching does not occur. 
In these plots the current required for the switching is compared with the Slonczewski's critical current $I_{sc}$.
Interestingly,  for $I_r'=5$, that corresponds to a very small nanomagnet comparable to a magnetic molecule, the critical current is higher than $I_{sc}$ if $\omega_0'<0.5$. However, above this value of $\omega_0'$ there is a region where the critical current required for the switching is lower than the Slonczewski's limit. At high resonance frequency of the oscillator the critical current coincides  with the Slonczewski's limit. For $I_r'=600$ (as well as for higher moments of inertia studied) the critical current also coincides with the Slonczewski's limit almost everywhere except for a narrow region of $\omega_0'$ between $0.5$ and $1$. Another interesting observation is the presence of sporadic non-switching islands down to the lowest values of $\omega_0'$ in the phase diagram $I/I_{sc}$ vs $\omega_0'$ for a very light oscillator. This observation can be of practical importance for current-induced magnetization switching in the smallest nanomagnets because it implies that some degree of the mechanical freedom may cause instability of the process.

\begin{figure}
\includegraphics[width=70mm]{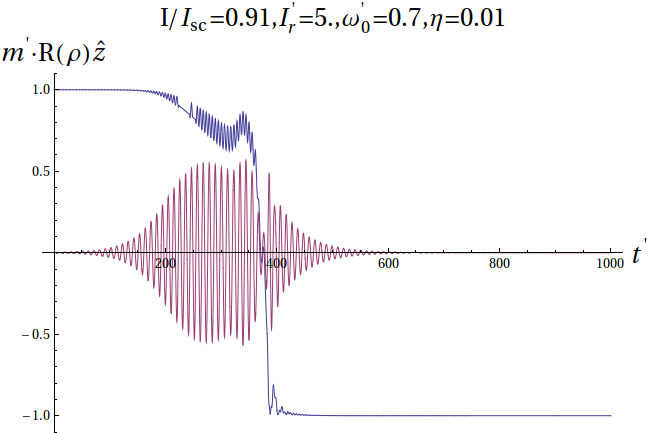}{a}
\includegraphics[width=70mm]{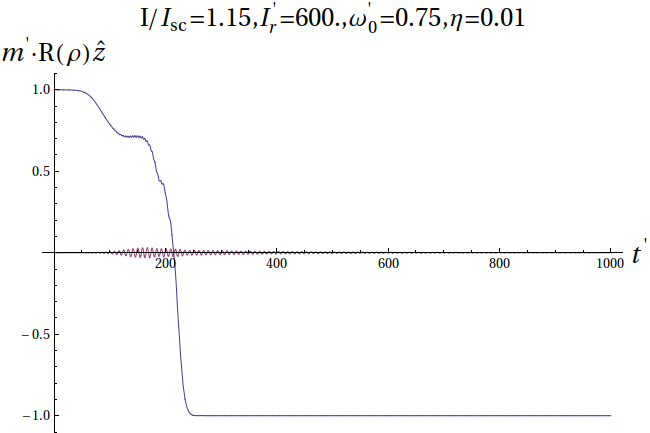}{b}
\includegraphics[width=70mm]{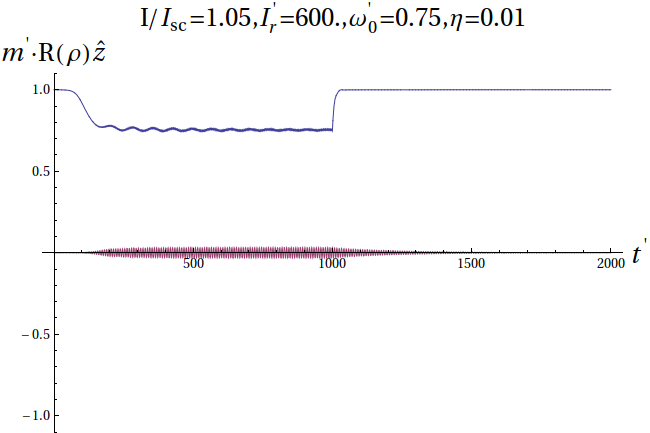}{c}
\caption{Typical current-driven switching (non-switching) dynamics of the magnetic moment in the coordinate frame of the oscillator, shown for three different sets of parameters.}
\label{cisoc}
\end{figure}

Three typical switching (non-switching) dynamics corresponding to particular points in the $I/I_{sc}$ vs $\omega_0'$ phase diagram are shown in Fig. \ref{cisoc}.
Comparison of switching speeds shows mixed results. In general, the switching speed in a small nanomagnet is lower than that in a large nanomagnet except when the current is close to the critical current or lower. This observation may appear counterintuitive. However, analysis of the equations shows that it is related to the fact that a heavier resonator absorbs the change in the spin angular momentum faster.

\section{Switching by the Current Combined with a Mechanical Kick} \label{combined}

\begin{figure}
\includegraphics[width=70mm]{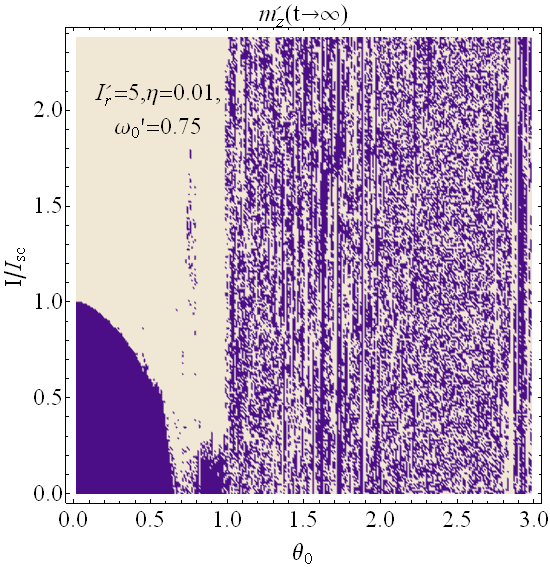}{a}
\includegraphics[width=70mm]{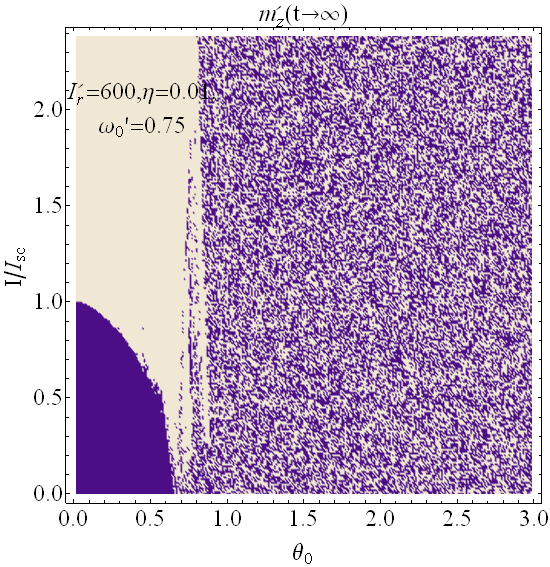}{b}
\caption{MAMS phase diagrams for current driven magnetization reversal accompanied by a mechanical kick.}
\label{PhaseIversustheta0}
\end{figure}
The analysis presented in the previous two Sections suggests that the maximum effect of the mechanical freedom of a nanomagnet may occur when the effect of the current is combined with a mechanical kick that twists the torsional oscillator by a significant angle. In this Section we solve equations  (\ref{dottheta}),  (\ref{dotphi}), and (\ref{rho-eq}) at a non-zero current, assuming that $\rho(t)$ is given by $\rho_0 cos(t)$ ($\theta_0 = \rho_0$) from $t'=0$ to $t'=T_i$ and by Eq.\ (\ref{rho-eq}) afterwards. Switching phase diagrams of $I/I_{sc}$ vs $\theta_0$ for $I_r'=5$ and $I_r'=600$ at $\eta=0.01$ and $\omega_0' = 0.75$ are shown in  Fig.\ \ref{PhaseIversustheta0}. 
There are strong similarities between cases of light and heavy oscillators, suggesting that the effect is quite universal. In agreement with previous results the switching occurs at $I > I_{sc}$ in the absence of the mechanical kick ($\theta_0 = 0$). Also in agreement with the results of Section \ref{mechanical} the switching occurs in certain windows of $\theta_0$ even at $I = 0$.  This observation may have practical importance because it shows the potential of MAMS for magnetization reversal by a lower spin-polarized current. As the moment of inertia increases, the width of the $I = 0$ switching region shrinks. At high $\theta_0$ the switching becomes sporadic as a consequence of the chaotic dynamics with uncertain outcome for $m_z$ at $t \rightarrow \infty$.

\section{Discussion} \label{discussion}

We have studied the effect of torsional oscillations of the magnet on the magnetization switching induced by the spin-polarized electric current. Our motivation for considering such a problem was two-fold. Firstly, applications of spin transfer torque require rather large currents. It is, therefore, natural to ask whether the process can be assisted by other means. Secondly, with various proposed designs of the magnetic memory operated by spin-polarized currents, it is not out of question that some magnetic elements would have a certain degree of mechanical freedom. The question we asked is whether mechanical vibrations inhibit or assist the magnetization reversal. Throughout the paper our focus has been on the mechanically assisted magnetization switching (MAMS) that can reduce the  minimal required current as compared to the Slonczewski's limit. 

In Section \ref{mechanical} we have shown that a mechanical kick alone, supplied to the oscillator, can switch the magnetization. In this case the effect of the mechanical oscillations is equivalent to the effect of the ac field. Indeed, the spins in the rotating coordinate frame of the resonator oscillating at an angular velocity $\dot{\rho}$ experience the effective ac magnetic field $h = \dot{\rho}/\gamma_g$. Weak ac magnetic fields are known to be capable of switching the magnetization through consecuitive absorption of photons. They transfer the angular momentum to the macrospin of the nanomagnet and drive it up the anisotropy barrier until the reversal occurs \cite{CGC-PRB2013}. As we have seen in Section \ref{mechanical}, torsional oscillations of the resonator have a similar effect on the magnetic moment. The  conditions required for the magnetization reversal are that the angle of twist is suffiiently large and that the resonance frequency of the mechanical oscillator is comparable to the frequency of the ferromagnetic resonance. The mechanical frequency of the torsional oscillator, $\omega_0 =\sqrt{k/I_r}$, depends on two parameters, the torsion elastic modulus $k$ and the moment of inertia $I_r$. The latter scales as the square of the oscillator size. Thus, the condition $\omega_0 \sim \omega_{FMR}$ requires nanoscale oscillators. Nowadays GHz nanomechanical oscillators are common. Oscillation frequencies of hundreds of GHz have been reported in carbon nanotubes \cite{Island-NL2012}. Even at lower frequencies, however, one observation made in Section \ref{current} may be important for some of the existing devices that use spin-transfer torque to achieve the magnetization reversal. We have seen in the switching phase diagram that any degree of the mechanical freedom of the device may lead to its instability due to the chaotic coupled dynamics of the magnetic moment and the mechanical twist inside certain windows of the values of the parameters. This effect becomes progressively weaker, however, as the oscillator becomes heavier. 

In the absence of the spin transfer torque delivered by the spin-polarized current, MAMS requires a large oscillation angle $\rho$, which can be viewed as a trade-off for a large current needed without the mechanical assistance. The necessity of a large initial twist for the MAMS unassisted by the current is easy to understand by noticing that the effective ac magnetic field in the coordinate frame of the oscillator is proportional to the amplitude of the oscillations, $h = \dot{\rho}/\gamma_g \sim \omega_0 \rho/\gamma_g$. In Section \ref{combined} we have shown that combining the effect of the spins-polarized current with the effect of high-frequency mechanical oscillations one can reduce both, the critical current and the amplitude of the oscillations required for the switching. This observation suggests a memory-switching nanodevice that is operated by both, spin-polarized current and electromechanical actuation. A fast mechanical  twist can be supplied to the oscillator through features used in the existing nanoelectromechanical systems (NEMS), by, e.g., placing it in the electric field and delivering a field pulse simultaneously with a pulse of the spin-polarized electric current. While this may be a challenging task at the nanoscale, the fast progress of nanotechnology leaves no doubt about its feasibility. The exact realization of the proposed combined switching mechanism may be far from the one schematically depicted in Fig. \ref{magnetic-molecule-model}. However, the equations derived and solved in this paper are based upon general physical principles and they should apply to a variety of situations in which the dynamics of the magnetization is coupled with the mechanical rotational motion.

\section{Acknowledgements}
This work has been supported by the U.S. National Science Foundation through Grant No. DMR-1161571.
Dr. Jaafar would like to acknowledge support from the PSC-CUNY Award, jointly funded by The Professional Staff Congress and The City University of New York. Numerical simulations carried under this research were supported, in part, under National Science Foundation Grants CNS-0958379 and CNS-0855217 and the City University of New York High Performance Computing Center at the College of Staten Island.

\end{document}